\documentclass[12pt]{article}
\usepackage{amssymb}
\usepackage[dvips]{graphicx}
\usepackage{amsmath,amscd}
\usepackage{color}
\usepackage[option]{colortbl}
\usepackage{subeqnarray}

\def\a{\alpha}
\def\a'{\alpha '}

\def\Lam{\Lambda}

\def\L{\mathcal{L}}

\def\Tr{{\rm Tr}}

\def\1{\amssymb{1}}

\def\lb {\left(}
\def\rb {\right)}

\def\llb {\left[}
\def\rlb {\right]}

\def\lan{\langle}
\def\ran{\rangle}

\newcommand {\beq}{\begin{eqnarray}}
\newcommand {\eeq}{\end{eqnarray}}
\newcommand {\ben}{\begin{enumerate}}
\newcommand {\een}{\end{enumerate}}
\newcommand {\bit}{\begin{itemize}}
\newcommand {\eit}{\end{itemize}}

\newcommand {\vphi}{\varphi}

\makeatletter
\@addtoreset{equation}{section}
\renewcommand{\theequation}{\thesection.\arabic{equation}}

\makeatother

\makeatletter
\newcommand{\thetablename}{Table}
\def\fnum@table{\thetablename\ \thetable}
\makeatother
\begin{document}

\thispagestyle{empty}
\begin{flushright}

\end{flushright}
\vspace{3mm}

\begin{center}
{\huge Inflaton versus Curvaton\\
  \vspace{2mm}in Higher Dimensional Gauge Theories}

\end{center}

\begin{center}
\lineskip .45em
\vskip1.5cm
{\large Takeo Inami\footnote{E-mail: inami@phys.chuo-u.ac.jp}$^1$,
Yoji Koyama\footnote{E-mail: koyama@phys.chuo-u.ac.jp}$^1$, Chia-Min Lin\footnote{E-mail: cmlin@phys.nthu.edu.tw}$^2$ and
Shie Minakami\footnote{E-mail: minakami@phys.chuo-u.ac.jp}}$^1$

\vskip 1.5em
{\large\itshape $^1$Department of Physics, Chuo University, Bunkyo-ku, Tokyo 112, Japan\\
\large\itshape $^2$Department of Physics, National Tsing Hua University, Hsinchu, Taiwan 300}  \vskip 4.5em
\end{center}

\begin{abstract}
We construct a model of cosmological inflation and perturbation based on the higher-dimensional gauge theory. The inflaton and curvaton are the scalar fields arising from the extra space components of the gauge field living in more than four dimensions. We take the six-dimensional (6D) Yang-Mills theory compactified on $T^2$ as a toy model, and apply the one-loop effective potential of the inflaton and the curvaton to the curvaton scenario. We have found that the curvaton is subdominant for the linear curvature perturbation, but that a significant non-Gaussianity and a sizable tensor to scalar ratio are generated.  
\end{abstract}


\newpage
\section{Introduction}
\indent \indent The mechanism of cosmological inflation and the way of generating the curvature perturbation are the two issues to be taken into account in constructing the particle physics model for the cosmological inflation. The former gives the proper initial condition for the Hot Big-Bang (HBB) model and the latter is the origin of the large-scale structure and the cosmic microwave background (CMB) anisotropies in the present Universe. The assumption that the inflaton (denoted by $\phi$) is responsible for both inflation and the curvature perturbation (inflaton senario) is very attractive. However the inflaton potential is subject to a few severe restrictions.\\
  \indent Lyth and Wands have put forward a new view that the curvature perturbation could 
be explained by another scalar field, called curvaton, different from the inflaton \cite{Lyth:2001nq,Enqvist:2001zp}. In this curvaton scenario, a non-Gaussian curvature perturbation may easily be generated, and the inflaton becomes free from the restriction from the curvature perturbation. It is now an important problem whether the inflaton scinario or inflaton plus curvaton scinario is suitable for explaining both inflation and the curvature perturbation. \\
  \indent The higher dimensional gauge theory with dimension $D$ larger than 5 has a room for two (or more) scalar fields. Hence this theory can be a good cosmological model to study the above problem of inflaton scenario versus inflaton plus curvaton scinario. The purpose of this work is to pursue the higher dimensional gauge theory as the model of such scalar field(s). To see whether the curvaton is really needed we should study both inflation and curvature perturbation simultaneously, unlike the previous studies. In the higher dimensional gauge theory a small and periodic (in $\phi$) potential arises through the radiative corrections. This picture was first applied to the Higgs potential \cite{Hatanaka:1998yp}, later to the inflaton potential \cite{ArkaniHamed:2003wu} and its supersymmetric extension is studied in \cite{Inami:2009bs}. It is also applied to the curvaton potential in \cite{Dimopoulos:2003az}. Applications of the periodic type potential to inflaton and curvaton potentials have been studied in the models of natural inflation \cite{Freese:1990rb} and hilltop inflation/curvaton \cite{Boubekeur:2005zm}. \\
 \indent  To show a possible realization of the curvaton model in the framework of the higher dimensional gauge theory, we take 6D $SU(2)$ gauge theory compactifed on $T^2$ as a toy model. The curvaton decay can be taken into account by adding a massless matter. Denoting the gauge field by $A_M$ $(M =0, 1, 2, 3, 5, 6)$, we identify the zero mode of its 5th and 6th components $A^{(0, 0)}_5$ and $A^{(0, 0)}_6$ with the inflaton $\phi$ and the curvaton $\sigma$, respectively. Here the suffix $(0, 0)$ means the zero modes with respect to the 5th and 6th coordinates $y_5$ and $y_6$.\\
  \indent Confronting the 6D model with all astrophysical data, we will find that the curvaton contribution to the linear part of the curvature perturbation is subdominant. However, the curvaton contribution could dominate the nonlinear part and detectable non-Gaussianity is generated. In a large area of the parameter space $f_{NL} \simeq 4$ and the spectral index $n_s=0.96$ (with negligible tensor to scalar ratio) are achieved. There is also some region in the parameter space with a detectable tensor to scalar ratio ($r \simeq 0.1$). Therefore our model can be constrained by the future observation \cite{Smith:2005mm}.

\section{One-Loop Effective Potential for the Inflaton and the Curvaton}

\indent \indent We consider 6D $SU(2)$ Yang-Mills theory compactified on $T^2$ with a massless matter $\psi$.  The Lagrangian is given by
\beq
\L=-\frac{1}{2}\Tr F_{MN}F^{MN}-i\bar{\psi}\gamma^MD_{M}\psi.
\eeq
The covariant derivative is defined as $D_M\psi \equiv (\partial_M-ig_6A_M)\psi$ where $g_{6}$ is the 6D gauge coupling constant.\\
\indent We compactify the 5th and 6th directions on $T^2$ with the radii $R_5$ and $R_6$. The gauge field $A_M(x^L)$ is then Fourier expanded in $y_5$ and $y_6$ as 
\beq
A_M(x^{\mu},y_5,y_6)=\frac{1}{\sqrt{2\pi R_5 2\pi R_6 }}\sum_{n,m=-\infty}^{\infty}A_M^{(n,m)}(x^{\mu})e^{i(ny_5/R_5+my_6/R_6)},\label{compact}
\eeq
same for $\psi$\footnote{It is possible to set a twisted boundary condition associated with a global $U(1)$ symmetry with the phase $\alpha$ for this fermion. Here we set $\alpha=0$.}. In (\ref{compact}) the zero modes $A^{(0,0)}_5$ and $A^{(0,0)}_6$ are 4D scalar fields and will be identified with the inflaton field $\phi$ and the curvaton field $\sigma$, respectively. \\
\indent Our central issue is the computation of the potential of $\phi$ and $\sigma$,$V(\phi,\sigma)$. To this end we allow $A_5^{(0,0)}$ and $A_6^{(0,0)}$ to have VEVs of the form 
\beq
\lan A_5^{(0,0)}\ran= \frac{1}{2\pi g_6R_5}\left(\begin{array}{cc}
\theta& 0\\
0 & -\theta
\end{array}\right),\quad
\lan A_6^{(0,0)}\ran= \frac{1}{2\pi g_6R_6}\left(\begin{array}{cc}
\vphi& 0\\
0 & -\vphi
\end{array}\right).
\eeq
$\theta$ and $\vphi$ are constants given by the Wilson line phases, $g_6\int_{0}^{2\pi R_a}dy_{a}\lan A_a^{(0,0)}\ran$ ($a=5,6$). \\
\indent At one-loop, the effective potential $V(\theta,\vphi)$ is obtained as the sum of the contributions of the gauge boson and the matter \cite{Antoniadis:2001cv}.
\beq
V(\theta,\vphi)&=&-\frac{R_5R_6}{\pi^7}\llb \sum_{k=1}^{\infty}\sum_{l=1}^{\infty}\frac{4}{(k^2R_5^2+l^2R_6^2)^3}\left((1+\cos (2k\theta))(1+\cos (2l\vphi))-2\cos (k\theta) \cos (l\vphi) \right)\right.\nonumber\\
&&\left. +\sum_{l=1}^{\infty}\frac{1}{(l^6R_6^6)}(1+\cos (2l\vphi)-2\cos (l\vphi))+\sum_{k=1}^{\infty}\frac{1}{(k^6R_5^6)}(1+\cos (2k\theta)-2\cos (k\theta))\rlb\nonumber \\
&& +\,{\rm const.}\label{potall}
\eeq
There is a divergent constant term corresponding to the cosmological constant. $V(\theta,\vphi)$ has minima at $(\theta,\vphi)=((2q+1)\pi,(2p+1)\pi)$ ($q,p$: integer).\\
\indent The leading terms ($k=1,l=1$) are a good approximation to (\ref{potall}) (see Figure 1).
\beq
V_{\rm eff}(\theta,\vphi)&\simeq&-\frac{R_5R_6}{\pi^7}\llb \frac{4}{(R_5^2+R_6^2)^3}\left((1+\cos (2\theta))(1+\cos (2\vphi))-2\cos \theta \cos \vphi \right)\right.\nonumber\\
&&\left. +\frac{1}{(R_6^6)}(1+\cos (2\vphi)-2\cos \vphi)+\frac{1}{(R_5^6)}(1+\cos (2\theta)-2\cos \theta)\rlb+{\rm const.}\nonumber \\ \label{appot}
\eeq

\begin{figure}[t]
\begin{center}
\begin{tabular}{ccc}
\begin{minipage}[b]{0.3\textwidth}
   \begin{center}
   \includegraphics[width=50mm]{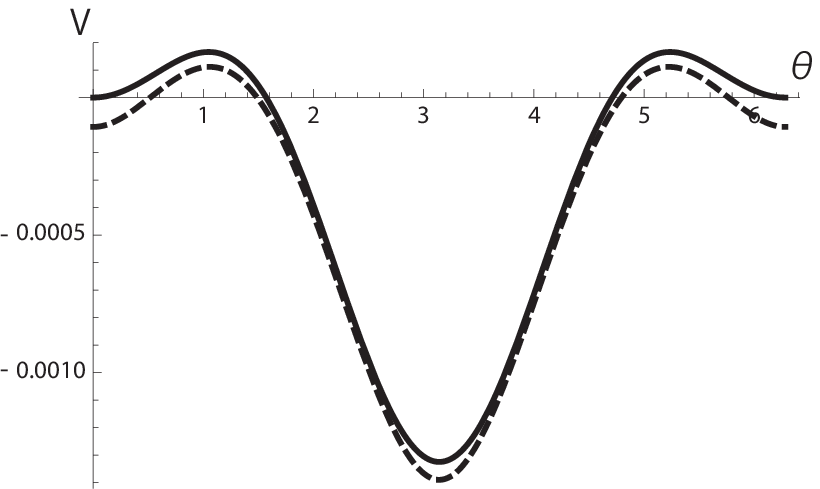}\\
   (a) 
 \end{center}
\end{minipage}  & 
\begin{minipage}[b]{0.3\textwidth}
 \begin{center}
   \includegraphics[width=50mm]{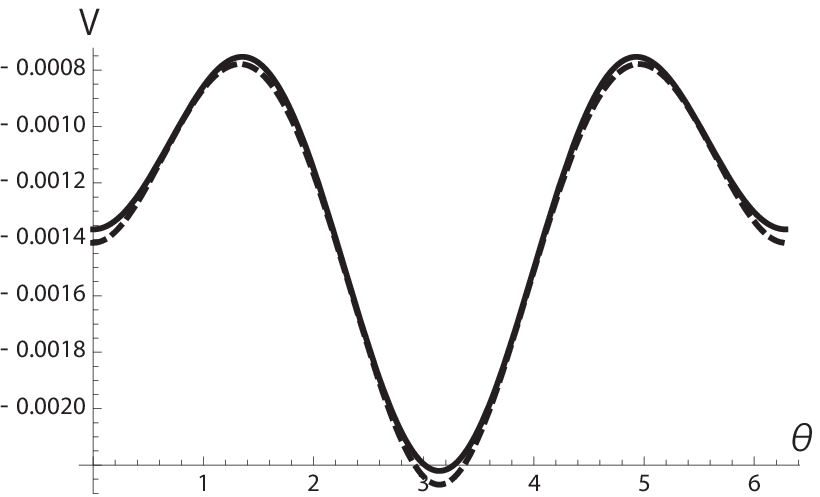}\\
   (b) 
  \end{center}
\end{minipage}  &
 \begin{minipage}[b]{0.3\textwidth}
  \begin{center}
   \includegraphics[width=50mm]{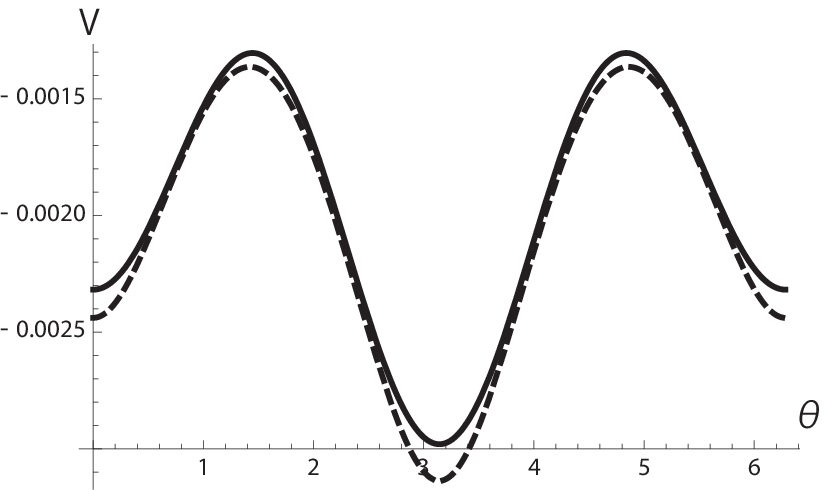}\\
   (c) 
 \end{center}
\end{minipage}   
\end{tabular}
\end{center}
\vskip-\lastskip
\caption{The potential $V(\theta,\vphi)$ with $R_5=R_6$ as a function of $\theta$ and for fixed values of $\vphi$. (a): $\vphi=\pi/2$. (b): $\vphi=3\pi/4$. (c): $\vphi=\pi$. The potential evaluated by taking the summation up to $k=100,l=100$ is represented by broken lines, the one by taking only the $k=1,l=1$ term by solid lines. The difference between the two is small. The unit of the vertical axes are $1/R_5^4$. The same convention is used in Figure 2.}
\label{fig:label-2}
\end{figure}
\indent We consider the theory around one of the potential minima and define the inflaton field $\phi$ and the curvaton field $\sigma$ as the fluctuations around it.
\beq
\phi \equiv f_5 (\theta(x^{\mu})-\pi), \quad \sigma \equiv  f_6 (\vphi(x^{\mu})-\pi),
\eeq
where
\beq
f_5=\frac{1}{2\pi g R_5}, \quad f_6=\frac{1}{2\pi g R_6}.
\eeq
$g=g_6/\sqrt{2\pi R_5 2\pi R_6}$ is the 4D gauge coupling constant. After the renormalization of the effective potential (\ref{appot}) so that it satisfies $V_{\rm eff}((2q+1)\pi,(2p+1)\pi)=0$,\footnote{This renormalization condition corresponds to the nearly zero cosmological constant $\Lam \simeq 10^{-12}\,({\rm eV})^4$ in our Universe.} and rewriting the potential in terms of $\phi$ and $\sigma$, we have
\beq
V(\phi,\sigma)&=&-\frac{R_5R_6}{\pi^7}\llb \frac{2}{(R_5^2+R_6^2)^3}\left((\cos (2\frac{\phi}{f_5})-1)(\cos (2\frac{\sigma}{f_6})-1)-2(\cos (\frac{\phi}{f_5})-1)(\cos (\frac{\sigma}{f_6})-1)\right)\right.\nonumber\\
&&\left. +\left(\frac{1}{R_6^6}+\frac{8}{(R_5^2+R_6^2)^3}\right)(\cos (2\frac{\sigma}{f_6})-1)+\left(\frac{2}{R_6^6}-\frac{8}{(R_5^2+R_6^2)^3}\right)(\cos (\frac{\sigma}{f_6})-1)\right.\nonumber\\
&&\left.+\left(\frac{1}{R_5^6}+\frac{8}{(R_5^2+R_6^2)^3}\right)(\cos (2\frac{\phi}{f_5})-1)+\left(\frac{2}{R_5^6}-\frac{8}{(R_5^2+R_6^2)^3}\right)(\cos (\frac{\phi}{f_5})-1)\rlb. \nonumber \\ \label{potic}
\eeq
The first line corresponds to the interaction terms between the inflaton and the curvaton, and the second and the third lines the self-interaction terms of the curvaton and those of the inflaton, respectively. We have three parameters, $g$, $R_5$ and $R_6$, in this theory. $V(\phi,\sigma)$ is shown in Figure 2, the contribution of the inflaton to the energy density of the Universe becomes dominant as the ratio $r_R\equiv R_5/R_6$ decreases.\\
\indent The derivatives of the potential will be necessary in the following sections and are defined by $V_{\phi}=\partial V(\phi,\sigma)/\partial \phi$ and $V_{\sigma}=\partial V(\phi,\sigma)/\partial \sigma$.
The inflaton mass ($m_{\phi}^2= V_{\phi \phi}|_{\phi=0}$) and curvaton mass are given by
\beq
m_{\phi}^2=\frac{4R_5R_6}{\pi^7f_5^2}\llb \frac{2}{(R_5^2+R_6^2)^3}\left(\cos (2\frac{\sigma}{f_6})-\frac12(\cos (\frac{\sigma}{f_6}))\right)+\frac{3}{2(R_5^2)^3}-\frac{5}{(R_5^2+R_6^2)^3}\rlb,\nonumber\\\label{minf}
\eeq
\beq
m_{\sigma}^2=\frac{4R_5R_6}{\pi^7f_6^2}\llb \frac{2}{(R_5^2+R_6^2)^3}\left(\cos (2\frac{\phi}{f_5})-\frac12\cos (\frac{\phi}{f_5})\right)+\frac{3}{2(R_6^2)^3}-\frac{5}{(R_5^2+R_6^2)^3}\rlb.\nonumber\\
\eeq
\begin{figure}[t]
\begin{center}
\begin{minipage}[b]{0.49\textwidth}
   \begin{center}
   \includegraphics[width=60mm]{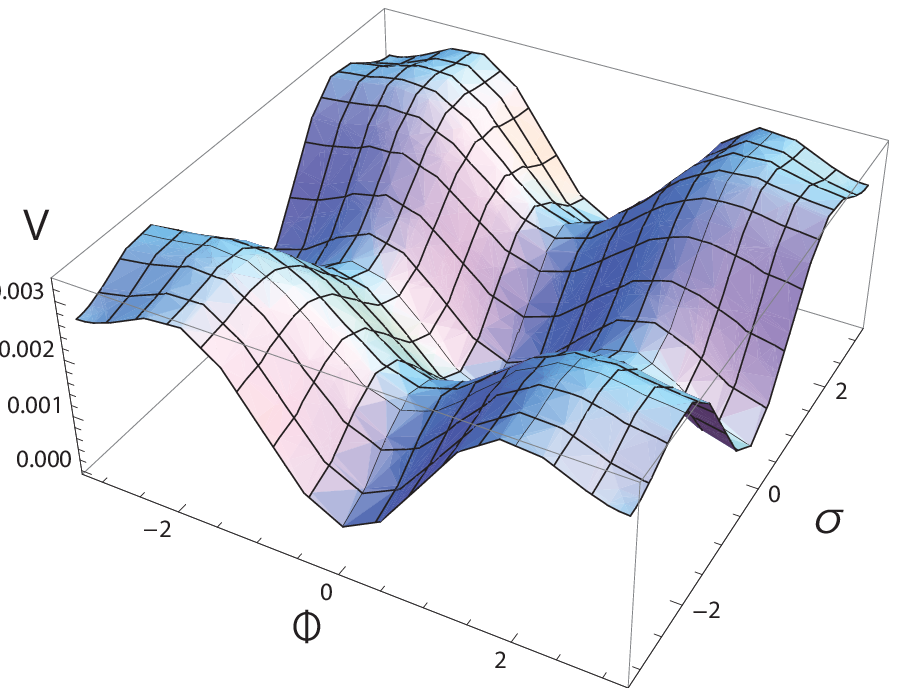}\\
   (d) 
   \end{center}
\end{minipage}   
 \begin{minipage}[b]{0.49\textwidth}
   \begin{center}
   \includegraphics[width=60mm]{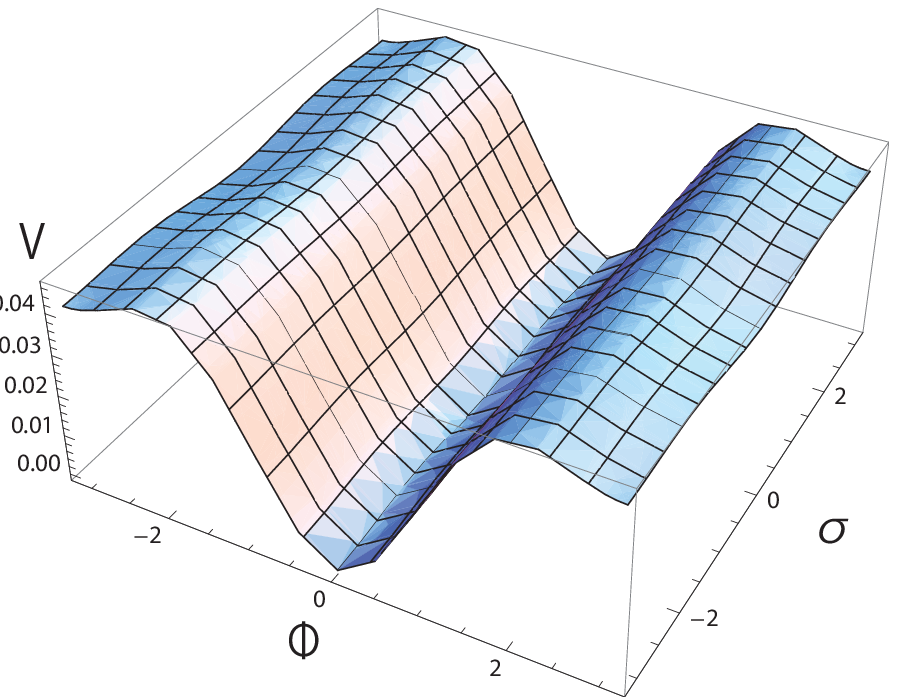}\\
   (e) 
   \end{center}
\end{minipage}   
\end{center}
\vskip-\lastskip
\caption{The effective potential $V(\phi,\sigma)$ with $R_5=r_RR_6(r_R\equiv R_5/R_6)$ for two values of $r_R$. (d): $r_R=1$.  (e): $r_R=0.5$.}
\label{fig:label-2}
\end{figure}

\section{Constraints for the Curvaton Model}

\indent \indent We apply the potential (\ref{potic}) to the curvaton scenario identifying the inflaton $\phi$ and the curvaton $\sigma$ with the higher-dimensional gauge fields $A^{(0,0)}_5$ and $A^{(0,0)}_6$, respectively. We deal with the power spectrum of the curvature perturbation ${\cal P}_{\zeta}$\footnote{ It is defined by the two point correlator of the curvature perturbation $\langle\zeta_{\vec k}\zeta_{\vec k'}\rangle=(16\pi^5/k^3)\delta^3_{{\vec k}+{\vec k'}}{\cal P}_{\zeta}(k)$ where $\zeta_{\vec k}$ is the Fourier component of the curvature perturbation. We will call ${\cal P}_{\zeta}$ simply the curvature perturbation. } due to the curvaton and that due to the inflaton, and denote them by ${\cal P}_{{\rm curv}}$ and ${\cal P}_{{\rm infl}}$. We consider the two alternative situations depending on the importance of ${\cal P}_{{\rm curv}}$ compared with ${\cal P}_{{\rm infl}}$, (I) ${\cal P}_{\zeta}\simeq{\cal P}_{{\rm curv}},\, \gg{\cal P}_{{\rm infl}}$ and (II) ${\cal P}_{\zeta}={\cal P}_{{\rm curv}}+ {\cal P}_{{\rm infl}}$.\\
\indent There are several aspects of inflation, theoretical and observational, which give constraints for the curvaton model \cite{Lyth:2001nq, Bartolo:2002vf,Komatsu:2010fb}, and will be taken into account in our analysis.
\begin{flushleft}
{\bf (I) ${\cal P}_{\zeta}\simeq{\cal P}_{{\rm curv}},\,\, {\cal P}_{{\rm infl}}\ll {\cal P}_{{\rm curv}}$}
\end{flushleft}\ben \vspace{-7mm}\item[(a)] The curvaton energy density should contribute negligibly during the inflation period in order to prevent the curvaton to contribute to inflation. This condition is met if
\beq
R_5<R_6.\label{a}
\eeq

\item[(b)] The curvaton should be nearly massless during the inflation period and in the state of constant energy. Quantitatively we should have
\beq
\left|\frac{m_{\sigma *}}{H_*}\right|\ll1,\label{b}
\eeq
where the star denotes the epoch of horizon exit and $m_{\sigma*}$ is the effective curvaton mass, $m_{\sigma*}^2=\partial^2 V(\phi,\sigma)/\partial \sigma^2|_{\phi_*, \sigma_*} = V_{\sigma \sigma}(\phi_*,\sigma_*)$. The Hubble parameter $H$ is related to the potential through the Friedman equation as
\beq
H^2=\frac{V(\phi,\sigma)}{3M_P^2},\label{H}
\eeq
where $M_P = 2.4\times10^{18}$ GeV is the {\it reduced} Planck mass. If $|m_{\sigma*}|\gtrsim|H_*|$, the curvaton begins to oscillate during the inflation period and it behaves as a matter and then dilute away due to inflation. The nearly masslessness condition (\ref{b}) is incompatible with parametric resonance, so the curvaton-type preheating does not occur \cite{Kohri:2009ac}. 

\item[(c)]  We require that inflation is not curvaton-driven. This is realized if the curvaton energy density $\rho_{\sigma}$ is still subdominant at the time when the curvaton oscillation begins (denoted by 'osc').
\beq
\rho_{\sigma}|_{{\rm osc}}\ll \rho_{\rm rad}|_{{\rm osc}}\simeq 3m^2_{\sigma}M^2_P,\label{c}
\eeq
$\rho_{\rm rad}$ is the energy density of the radiation due to the inflaton decay. The curvaton begins to oscillate after the inflaton decay (reheating). We note that after reheating $\phi$ particles are no more present and the background field $\phi$ takes the VEV, $\lan \phi \ran=\phi_0$. In our model, $\phi_0=0$, hence the potential takes the form,
\beq
V(\phi,\sigma)|_{\rm after\,\, reheating}=V(0,\sigma),
\eeq
\beq
\rho_{\sigma}|_{\rm osc}=V(0,\sigma_*).\label{rhos}
\eeq 
 
\item[(d)] The slow-roll conditions
\begin{subeqnarray}
\epsilon &\equiv& \frac{1}{2}M_{P}^2\frac{V_{\phi}^2+V_{\sigma}^2}{V(\phi,\sigma)^2}\simeq \frac{1}{2}M_{P}^2\lb\frac{V_{\phi}}{V(\phi,\sigma)}\rb^2\ll1,\label{eps}\\ 
|\eta_{\phi \phi}| &\equiv& M_{P}^2\left|\frac{V_{\phi \phi}}{V(\phi,\sigma)}\right|\ll1,\label{d}
\end{subeqnarray}
 The second equality in (3.7a) holds true under the inflaton dominance of the energy density. Then it is obvious that $V_{\phi}\gg V_{\sigma}$ from Fig. 2 (e). Note that (3.7a) is equivalent to the usual definition of $\epsilon$, $\epsilon\equiv -\dot{H}/H^2$.

 \item[(e)] For the flatness problem and the horizon problem to be solved, the number of e-foldings $N$ has to be at least 50-60.
\beq
N \simeq \frac{1}{M_P^2} \left| \int_{\phi_*}^{\phi_f} \frac{V(\phi,\sigma)}{V_{\phi}} \,d\phi \right|\simeq50-60.\label{e}
\eeq
$\phi_f$ is the values of $\phi$ at the time when the slow-roll condition ends, namely when $\epsilon$ and $\eta$ become nearly 1. 

\item[(f)] The curvaton perturvation should be nearly Gaussian in accordance with the current observations.
\beq
\sigma^2_* \gg\delta\sigma^2=\left(\frac{H_*}{2\pi}\right)^2.\label{f}
\eeq

\item[(g)] ${\cal P}_{\rm curv}$ must reproduce the observed anisotropies in CMB \cite{Komatsu:2010fb}. 
\beq
{\cal P}_{\rm curv}=\frac49 \Omega^2 \left(\frac{g'(\sigma_*)}{g(\sigma_*)}\right)^2 \left(\frac{H_*}{2\pi}\right)^2 =  2.45 \times 10^{-9},\label{g}
\eeq
where \cite{Bartolo:2002vf}
\beq
\Omega\equiv \left(\frac{\rho_{\sigma}}{\rho_{\rm rad}}\right)_{\rm dec}=\left(\frac{\rho_{\sigma}}{\rho_{\rm rad}}\right)_{\rm osc}\left(\frac{m_{\sigma}}{{\it \Gamma}_{\sigma}}\right)^{1/2}, \quad g(\sigma_*)=2\frac{\rho_{\sigma}|_{\rm osc}}{m^2_{\sigma}}.\label{omega}
\eeq
The prime in (\ref{g}) denotes the derivative with respect to $\sigma$. 'dec' denotes the time of the curvaton decay. The curvaton decays into gauge bosons and matter with the decay rate 
\beq
{\it \Gamma}_{\sigma}=\frac{g^2}{4\pi}m_{\sigma}.\label{gamma}
\eeq
\item[(h)] ${\cal P}_{\rm infl}$ is assumed to be small,
\beq
{\cal P}_{\rm infl}=\frac{1}{24\pi^2 M^4_P}\frac{V(\phi_*,\sigma_*)}{\epsilon_*}\ll10^{-9}.\label{h}
\eeq

\item[(i)] The $f_{NL}$ for non-Gaussianity in the curvaton
scenario is given by \cite{Lyth:2005du}
\beq
\frac35 f_{NL}=-1-\frac23 \Omega+\frac34\Omega^{-1}\left(1+\frac{g''(\sigma_*)g(\sigma_*)}{{g'}^2(\sigma_*)}\right).
\eeq
The recent observation indicates
\beq
f_{NL}<100.\label{i}
\eeq 
  
\item[(j)] The spectral index $n_s$ is determined from the WMAP deta \cite{Komatsu:2010fb}.
\beq
n_s \equiv 1-2\epsilon_*+2\eta_{\sigma \sigma*}= 0.96,\quad \eta_{\sigma \sigma}\equiv M_{P}^2\frac{V_{\sigma\sigma}(\phi,\sigma)}{V(\phi,\sigma)}.\label{j}
\eeq

\item[(k)] There is an upper bound for the tensor to scalar ratio $r$ from the observations \cite{Komatsu:2010fb}.
\beq
r=\frac{{\cal P}_h}{{\cal P}_{\zeta}}\lesssim0.2,\label{l}
\eeq
where ${\cal P}_h$ is the spectrum of the primordial tensor perturbation.  
\beq
{\cal P}_h=\frac{8}{M^2_P} \left(\frac{H_*}{2\pi}\right)^2.
\eeq

\een

\begin{flushleft}
{\bf (II) ${\cal P}_{\zeta}={\cal P}_{{\rm curv}}+ {\cal P}_{{\rm infl}}$}
\end{flushleft}\vspace{-5mm}In this case the conditions are the same as those of the case (I) except for (g) through (j). These four should be replaced by the following three \cite{Ichikawa:2008iq}.
\ben

\item[(g')] Both the curvaton and the inflaton contribute to the curvature perturbation,
\beq
{\cal P}_{\zeta}={\cal P}_{{\rm curv}}+ {\cal P}_{{\rm infl}}=2.45 \times 10^{-9}.\label{g'}
\eeq

\item[(i')] The $f_{NL}$ for the non-Gaussianity
\beq
f_{NL}&=&\frac{5}{6\left(1+\frac89\epsilon_*M^2_P \Omega^2 \left(\frac{g'(\sigma_*)}{g(\sigma_*)}\right)^2\right)^2}\left[2\epsilon_*-\eta_*+\frac{32}{27}\epsilon^2_*M^4_P \Omega^2 \left(\frac{g'(\sigma_*)}{g(\sigma_*)}\right)^2\frac{\partial}{\partial \sigma}\left(\Omega \frac{g'(\sigma_*)}{g(\sigma_*)}\right)\right] \nonumber \\ 
&&<100.\label{i'}
\eeq

\item[(j')] The spectral index $n_s$
\beq
n_s=1-2\epsilon_*-\frac{4\epsilon_*-2\eta_*}{1+\frac89\epsilon_*M^2_P \Omega^2 \left(\frac{g'(\sigma_*)}{g(\sigma_*)}\right)^2}=0.96.\label{j'}
\eeq

\een
There is the condition that the curvaton need to decay before the BB nucleosynthesis, ${\it \Gamma}_{\sigma}>H_{BBN}\simeq6.0\times 10^{-12}{\rm eV}$, but it does not give a significant restriction on the model.  

\section{Numerical Study}

\indent \indent  We proceed to study the question whether our cosmological inflation model can reproduce the astrophysical data by considering the conditions discussed in the last section.  We use the set of parameters \{$R_5, r_R\equiv R_5/R_6, f_5=1/(2\pi g R_5), \phi_*, \sigma_*$\}. In terms of $R_5$, $r_R$ and $f_5$, the potential (\ref{potic}) is written as
\beq
V(\phi,\sigma)&=&-\frac{1}{\pi^7 r_R R_5^4}\nonumber \\
&& \times \llb \frac{2}{(1+r_R^{-2})^3}\left((\cos (2\frac{\phi}{f_5})-1)(\cos (2\frac{\sigma}{r_R f_5})-1)-2(\cos (\frac{\phi}{f_5})-1)(\cos (\frac{\sigma}{r_R f_5})-1)\right)\right.\nonumber\\
&&\left. +\left(r_R^6+\frac{8}{(1+r_R^{-2})^3}\right)(\cos (2\frac{\sigma}{r_R f_5})-1)+\left(2r_R^6-\frac{8}{(1+r_R^{-2})^3}\right)(\cos (\frac{\sigma}{r_R f_5})-1)\right.\nonumber\\
&&\left.+\left(1+\frac{8}{(1+r_R^{-2})^3}\right)(\cos (2\frac{\phi}{f_5})-1)+\left(2-\frac{8}{(1+r_R^{-2})^3}\right)(\cos (\frac{\phi}{f_5})-1)\rlb. 
\eeq
\indent The above question is equivalent to asking whether there are parameter regions where all of the conditions are satisfied. First we consider the realization of the slow-roll inflation.\\
\indent 1) Determine $f_5$ and $\phi_*$ so that the conditions (d) and (e) are satisfied. The condition that the curvaton contribution to the energy density is negligible gives rise to an upper bound for $r_R$. After the procedure 1), the condition (b) is satisfied automatically by (\ref{b}), (\ref{H}) and $\eta_{\phi \phi}>\eta_{\sigma \sigma}$. The remaining three parameters $R_5$, $r_R$ and $\sigma_*$ are determined by considering the conditions for ${\cal P}_{\zeta}$ and $n_s$, eqs. (\ref{g}) and (\ref{j}) in the case (I) and eqs. (\ref{g'}) and (\ref{j'}) in the case (II).\\
\indent 2) Solve the equation (\ref{j}) ((\ref{j'})) with respect to $R_5=R_5(r_R, \sigma_*)$ and plot ${\cal P}_{\zeta}(r_R,\sigma_*)=2.45\times 10^{-9}$ on the $r_R-\sigma_*$ plane.\\
\indent 3) Finally we have to check that such $r_R$ and $\sigma_*$ are suitable for the curvaton model. To this end we evaluate the remaining conditions, inequalities (\ref{c}), (\ref{f}), (\ref{h}), (\ref{i}) and (\ref{l}) for the case (I), and (\ref{c}), (\ref{f}), (\ref{i'}) and (\ref{l}) for the case (II) to estimate the values of the parameters $r_R$, $\sigma_*$ and hence $R_5$. 
\begin{flushleft}
{\bf (I) ${\cal P}_{\zeta}\simeq{\cal P}_{{\rm curv}},\,\, {\cal P}_{{\rm infl}}\ll {\cal P}_{{\rm curv}}$}
\end{flushleft} \vspace{-5mm}
\indent \indent 1) For the slow-roll inflation, (d) and (e) are met if $f_5\gtrsim10M_P$. We take two typical values of $f_5$, A) $10M_P$ and B) $10^2M_P$. \\
A) $f_5=10M_P$. We can neglect the contribution of the curvaton to the energy density for $r_R\lesssim0.2$ which is the condition that the variation of $\sigma$ does not produce visible changes of the value of the potential and its derivatives ($\epsilon$ and $\eta_{\phi \phi}$). The condition (e) leads to $\phi_*=13M_P$ for $N\simeq55$.\\
\indent 2) We have tried to solve (\ref{j}) for $\phi_*=13M_P$, but there is no solution to $n_s(R_5,r_R,\sigma_*)=0.96$ (we get $n_s=0.98-0.99$). It is widely known that the curvaton dominance tends to give a larger value of $n_s$ ($\sim 0.98$) compared with the precise data $n_s=0.963\pm 0.012$. In our model the curvaton dominance does not holds, unless we allow an artificially larger error to $n_s$.\\ 
B) $f_5=10^2M_P$. The condition (a) and (e) lead to $r_R\lesssim0.1$ and $\phi_*=15M_P$ for $N\simeq55$. \\
\indent 2) We obtain a similar result as the case A) for $n_s$, and the same result holds for larger values of $f_5$.\\
\indent Thus we conclude that the case (I) of curvaton dominance in the curvature perturbation is incompatible with the curvaton model based on the higher-dimensional gauge theory.
 \begin{flushleft}
{\bf (II) ${\cal P}_{\zeta}={\cal P}_{{\rm curv}}+ {\cal P}_{{\rm infl}}$}
\end{flushleft} \vspace{-5mm}
\indent \indent 1) For the slow-roll inflation, (d) and (e) are met if $f_5\gtrsim10M_P$.\\
A) $f_5=10M_P$. The conditions (a), (d) and (e) are the same as the case (I), so we have the same results for $r_R$ and $\phi_*$, $r_R\lesssim 0.2$ and $\phi_*=13M_P$ for $N\simeq55$.\\
\indent 2) We consider the curvature perturbation (\ref{g'}) and the spectral index (\ref{j'}) as functions of $r_R$, $\sigma$ and $R_5$. Under the relations (\ref{H}), (\ref{rhos}), (\ref{omega}) and (\ref{gamma}) they are written as
\beq
{\cal P}_{\zeta}&=&{\cal P}_{{\rm infl}}+ {\cal P}_{{\rm curv}}\nonumber \\
&=& \frac{V(\phi_*,\sigma_*)}{M_P^2}\left(\frac{1}{24\pi^2M_P^2\epsilon_*}+\frac{16\pi R_5^2Q_{\sigma}^2(\sigma_*)}{27}\right)=2.45\times10^{-9},\label{P}
\eeq
and
\beq
n_s=1-2\epsilon_*-\frac{4\epsilon_*-2\eta_*}{1+\frac{128\pi^3}{9}\epsilon_*M^2_PR_5^2Q_{\sigma}^2(\sigma_*)}=0.96,\label{n}
\eeq
where
\beq
Q_{\sigma}(\sigma_*)\equiv f_5\frac{V_{\sigma}(0,\sigma_*)}{3m_{\sigma}^2M_P^2}.
\eeq
We can easily solve (\ref{n}) with respect to $R_5$ and have
\beq
R_5^2=\left(\frac{4\epsilon_*-\eta_*}{0.04-2\epsilon_*}-1\right)\left(\frac{128\pi^3}{9}\epsilon_*M_P^2Q_{\sigma}^2(\sigma_*)\right)^{-1}.\label{R}
\eeq
Substituting (\ref{R}) to (\ref{P}), we obtain ${\cal P}_{\zeta}$ as a function of $r_R$ and $\sigma$, ${\cal P}_{\zeta}(r_R,\sigma_*)=2.45\times10^{-9}$. Its solution curve is shown in Figure 3 (1). 
\begin{figure}[t]
\begin{center}
\begin{minipage}[b]{0.49\textwidth}
   \begin{center}
   \includegraphics[width=65mm]{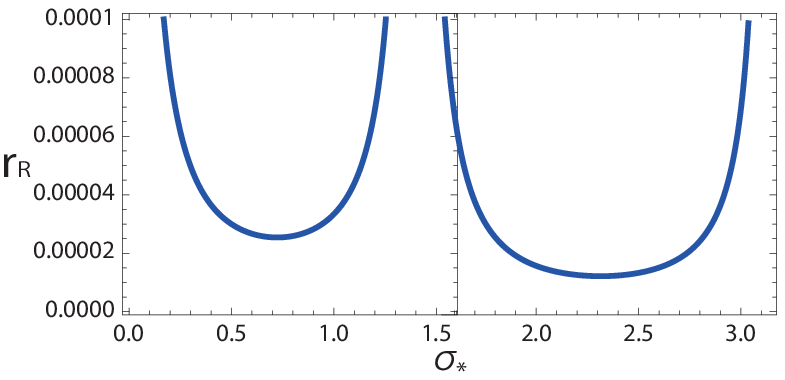}\\ 
   (1)
   \end{center}
\end{minipage}   
\begin{minipage}[b]{0.49\textwidth}
   \begin{center}
   \includegraphics[width=65mm]{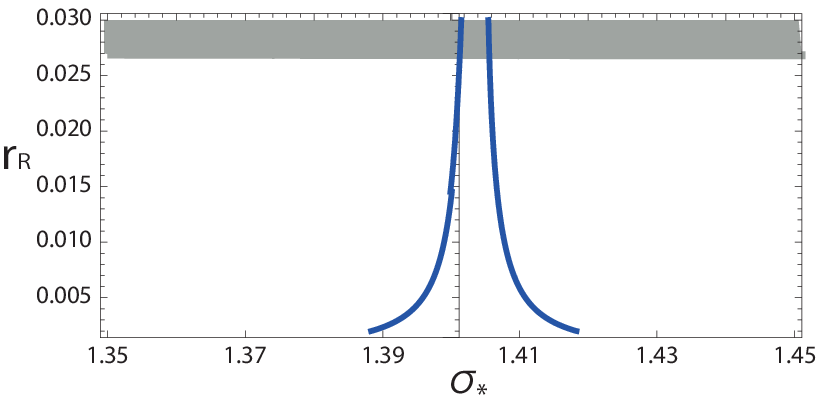}\\
   (2)
   \end{center}
\end{minipage}   
\begin{minipage}[b]{0.49\textwidth}
   \begin{center}
   \includegraphics[width=65mm]{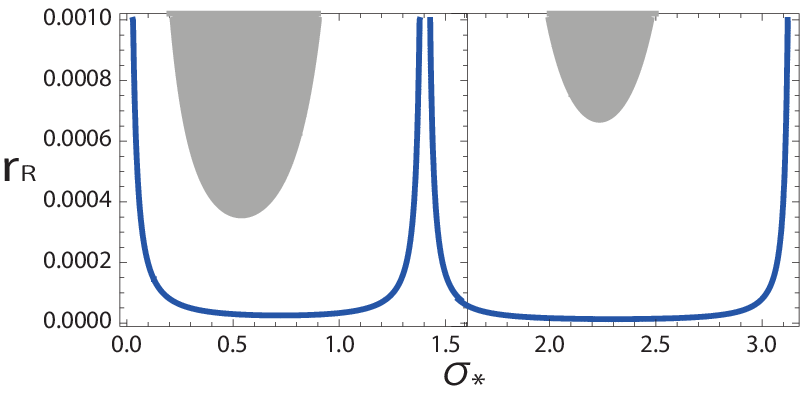}\\
   (3) 
   \end{center}
\end{minipage}  
 \begin{minipage}[b]{0.49\textwidth}
   \begin{center}
   \includegraphics[width=65mm]{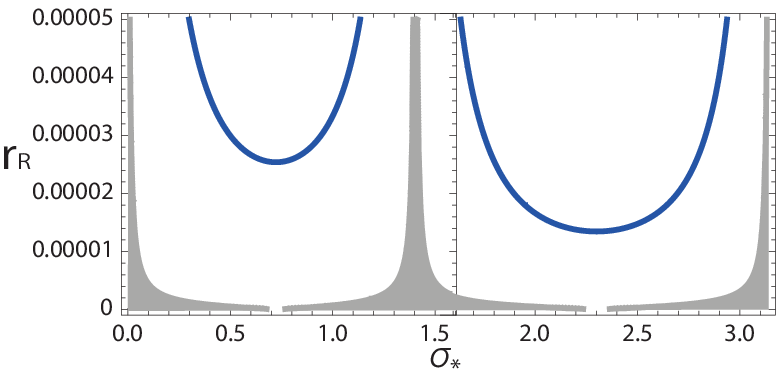}\\
   (4)
   \end{center}
\end{minipage}   
 
\begin{minipage}[b]{0.49\textwidth}
   \begin{center}
   \includegraphics[width=65mm]{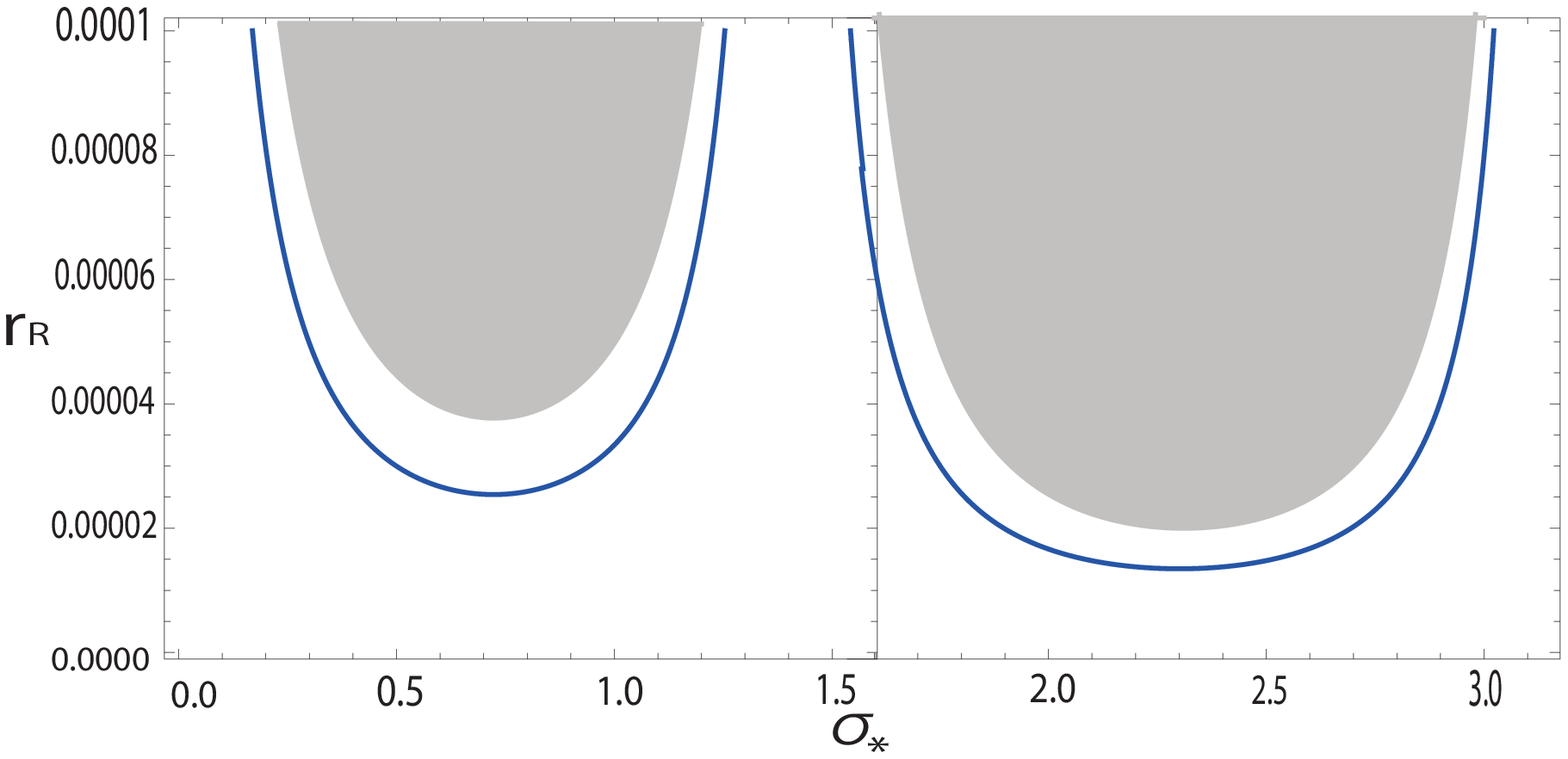}\\
   (5) 
   \end{center}
\end{minipage}
 \end{center}
\vskip-\lastskip
\caption{The line of constant ${\cal P}_{\zeta}$ in the $r_R-\sigma$ plane, the blue line representing ${\cal P}_{\zeta}(r_R,\sigma_*)=2.45\times10^{-9}$ for $f_5=10M_P$ and $\phi_*=13M_P$ together with the inequalities in the conditions (c), (f), (i') and (k). The shaded areas in (2), (3), (4) and (5) represent the regions forbidden by the inequalities (\ref{c}), (\ref{f}), (\ref{i'}) and (\ref{l}), respectively. $\sigma_*$ is in the unit of $r_Rf_5$. The same convention is used in Figure 4.}
\label{fig:label-2}
\end{figure}
The ratio of ${\cal P}_{\rm infl}$ to ${\cal P}_{\rm curv}$ turns out to be almost independent of the values of $r_R$ and $\sigma_*$ and is   
\beq
\frac{{\cal P}_{\rm curv}}{{\cal P}_{\rm infl}}\simeq 0.04.\label{ratio}
\eeq
Hence the curvaton contribution to the curvature perturbation is not sizable.\\
\indent 3) We evaluate the remaining conditions, (c), (f), (i') and (k). To evaluate the inequalities in (\ref{c}) and (\ref{f}) numerically, we replace it to as follows, 
\beq 
\rho_{\sigma}|_{{\rm osc}}\ll 3m^2_{\sigma}M^2_P\ \rightarrow \ \rho_{\sigma}|_{{\rm osc}}< \frac{3m^2_{\sigma}M^2_P}{100}, \quad {\rm for\,\, (\ref{c})},\nonumber
\eeq
and
\beq
\sigma^2_* \gg \left(\frac{H_*}{2\pi}\right)^2\ \rightarrow \ \frac{\sigma^2_*}{100} >\left(\frac{H_*}{2\pi}\right)^2, \quad {\rm for \,\, (\ref{f})}.\nonumber
\eeq
In Figure 3, we see that all of the conditions except for the condition (c) don't give additional restrictions to the values of $r_R$ and $\sigma_*$. The condition (c) yields the bound on $\sigma_*$, $ \sigma_* \lesssim 9.1 \times 10^{17}\,{\rm GeV}$. As a result, we have 
\beq
3.4\times10^{14}\,{\rm GeV}\lesssim \sigma_*\lesssim 9.1 \times 10^{17}\,{\rm GeV},\quad 2.4\times10^{-5}\lesssim r_R\lesssim0.2.\label{sr}
\eeq
 The $\phi_*$ is fixed from the value of $N$, (\ref{e}), whereas the $\sigma_*$ has a range coupled with $r_R$. \\
\begin{table}[tbp]
\newcommand{\lw}[1]{\smash{\lower0.3ex\hbox{#1}}}
\begin{center}
\renewcommand{\arraystretch}{1.2}
\begin{tabular}{|c|c|c|c|c|c|c|c|}
\hline
 \lw{$\sigma_*[{\rm GeV}]$}&\lw{$r_R$}&\lw{$g$}&\lw{$R_5$ $[{\rm GeV}^{-1}]$}&\lw{$R_6$ $[{\rm GeV}^{-1}]$}&\lw{$f_{NL}$}&\lw{$r$}\\
\noalign{\hrule height 0.6pt}
$9.1 \times 10^{17}$&$2.7\times10^{-2}$ &$1.7\times10^{-4}$&$3.9\times10^{-17}$&$1.4\times10^{-15}$ &$1.2$&$0.07$\\
$2.6\times10^{15}$&$0.20$ &$3.8\times10^{-4}$&$1.7\times10^{-17}$&$8.5\times10^{-17}$ &$0.5$&$0.06$\\
$1.3\times10^{15}$&$2.0\times10^{-5}$ &$4.0\times10^{-5}$&$1.7\times10^{-16}$&$8.3\times10^{-12}$ &$4.3$&$0.06$\\
$8.4\times10^{14}$&$1.6\times10^{-5}$ &$4.2\times10^{-5}$&$1.6\times10^{-16}$&$9.8\times10^{-12}$ &$1.1$&$0.1$\\

\hline
\end{tabular}
\end{center}
\caption{Sample sets of the parameters.}
\label{table}
\end{table}%
\indent Now we have all the parameters of our model. Sample sets of the parameters and the corresponding values of $f_{NL}$ and $r$ are shown in Table 1. We have the parameters of the theory as
 \beq
 g \simeq (3.8\,-\,0.4)\times10^{-4},\qquad \nonumber \\
 R_5\simeq (1.7\,-\,0.17)\times10^{-16}\, \mbox{GeV}^{-1},\nonumber \\
 R_6\simeq (9.8\times10^{-12}\,-\,8.5\times10^{-17})\, \mbox{GeV}^{-1}.\label{g}
\eeq
Due to the requirements of the slow-roll inflation and the tiny curvature perturbation, the smallness of the  gauge coupling constant $g$ is unavoidable in our model. The situation may change drastically, if we introduce supersymmetry (SUSY) and its breaking. In our previous study of the inflaton model from 5D SUSY gauge theory, the coupling constant $g$ is as large as $0.2$ \cite{Inami:2009bs}\footnote{In our previous paper \cite{Inami:2009bs} we made an error in estimating the coupling constant $g$. The value $g\lesssim 0.85$ should read $g\lesssim 0.2$.}. The 5th compactification radius $R_5$ which is related to the inflaton through $f_5$ is also required to be small by the slow-roll condition. However it is still larger than the Planck length $1/M_P$ at which the quantum gravity effects become dominant. $R_6$ is larger than $R_5$ by an order ${\cal O}(10)$ to ${\cal O}(10^5)$.\\
\indent The non-Gaussianity parameter $f_{NL}$ and the tensor-scalar ratio $r$ are in the range, 
\beq
 f_{NL}\simeq 4.3\ - \ 0.05,\nonumber \\
 r\simeq 0.1\ -\ 0.05. 
\eeq
 We saw that the linear part of the curvature perturbation is mostly due to the inflaton, (\ref{ratio}). But thanks to the curvaton, unlike the single field inflation model in which $f_{NL}$ is of order the slow-roll parameter, $f_{NL}$ is as large as $4$. The tensor-scalar ratio is given by $r\simeq16\epsilon_*$ as the case of the single field inflation. In this case, the inflaton mass $m_{\phi}$ is fixed by the constraint of the curvature perturbation and is given by $m_{\phi}\sim10^{13}\,{\rm GeV}$ \cite{Inami:2009bs}.\\      
B) $f_5=10^2M_P$. The same argument as the case A) applies to this case. We find from Figure 4 that there is no region of $r_R$ and $\sigma_*$ satisfying the conditions (c), (f), (i') and (k) simultaneously. The same result holds for larger values of $f_5$. We conclude that the case of $f_5\gtrsim10^2M_P$ is unsuitable for the curvaton model.
\begin{figure}[tbp]  
\begin{center}
\begin{minipage}[b]{0.49\textwidth}
   \begin{center}
   \includegraphics[width=65mm]{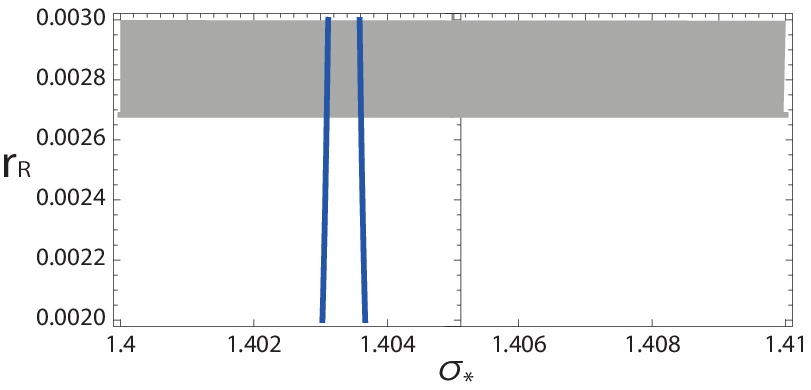}\\
   (1)
   \end{center}
\end{minipage}   
\begin{minipage}[b]{0.49\textwidth}
   \begin{center}
   \includegraphics[width=65mm]{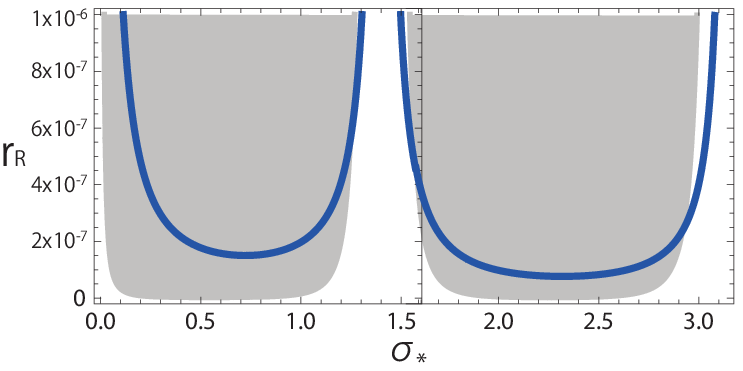}\\
   (2) 
   \end{center}
\end{minipage}  
 \begin{minipage}[b]{0.49\textwidth}
   \begin{center}
   \includegraphics[width=65mm]{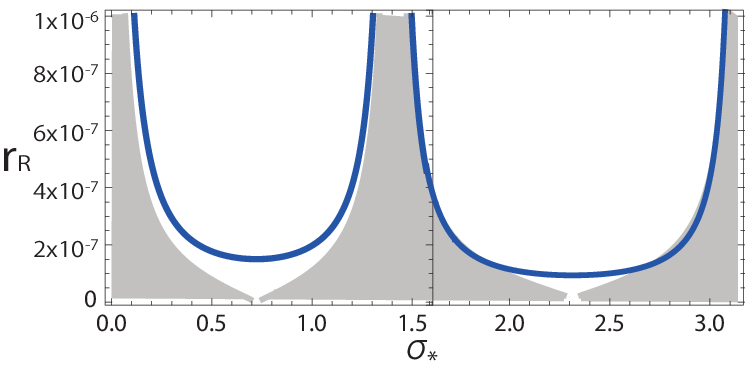}\\
   (3)
   \end{center}
\end{minipage}   
\begin{minipage}[b]{0.49\textwidth}
   \begin{center}
   \includegraphics[width=65mm]{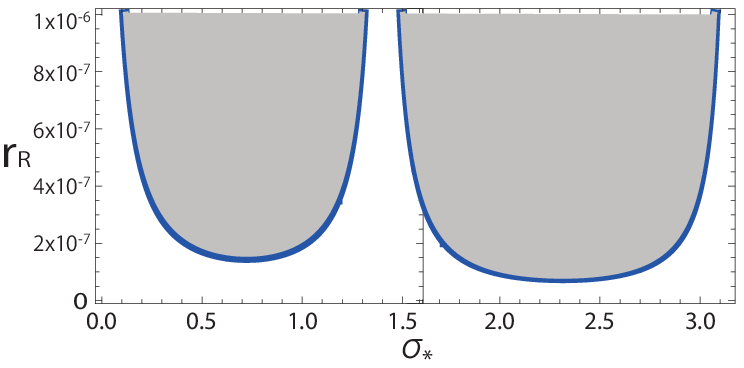}\\
   (4) 
   \end{center}
\end{minipage}
 \end{center}
\vskip-\lastskip
\caption{The line of constant ${\cal P}_{\zeta}$ in the $r_R-\sigma$ plane, the blue line representing ${\cal P}_{\zeta}(r_R,\sigma_*)=2.45\times10^{-9}$ for $f_5=10^2M_P$ and $\phi_*=14M_P$ together with the inequalities in the conditions (c), (f), (i') and (k). The sheded areas in (1), (2), (3) and (4) represent the regions forbidden by the inequalities (\ref{c}), (\ref{f}), (\ref{i'}) and (\ref{l}), respectively. }
\label{fig:label-2}
\end{figure}
\section{Discussion}
\indent \indent We have applied the 6D $SU(2)$ gauge theory to the inflaton-curvaton model identifying $A_5^{(0,0)}$ and $A_6^{(0,0)}$ with the inflaton $\phi$ and the curvaton $\sigma$. We have found that this model can explain the astrophysical data {\em only when} it is applied to the case A) in the situation (II), both the inflaton and the curvaton contributing to the curvature perturbation. But the contribution of the curvaton to the linear part of the curvature perturbation ${\cal P}_{{\rm curv}}$ is small compared with the inflaton part ${\cal P}_{{\rm infl}}$. Hence the curvaton is only responsible for generating the non-Gaussian perturbation. The non-Gaussianity parameter will soon be measured to the accuracy of $\delta f_{NL}\simeq 1$ \cite{Komatsu:2009kd}. $f_{NL}\simeq4$ (see Table 1) has a good chance of detection. The tensor to scalar ratio $r$ is in the range $r\simeq0.1\, - \,0.05$ in our model. This value is also large enough to be detected in the near future measurements of $r$. PLANCK, QUIET, PolarBear and DECIGO \cite{Komatsu:2009kd,Smith:2005mm} plan to detect $r$ with the accuracy of $r\simeq0.01$. Note in Table 1 that there is a large parameter region in which both $f_{\rm NL}$ and $r$ take large detectable values. \\
\indent It remains to be an open question whether there are means to remedy the smallness of the gauge coupling constant $g$ of our toy model compared with the realistic GUT values (by an order ${\cal O}(10^{-3})$). One possible way may be to construct supersymmetric $6D$ gauge theory, and it will be our future work to apply it to the inflaton and curvaton. \\
\indent In this work we have concentrated on the role of the zero modes $A_5^{(0,0)}$ and $A_6^{(0,0)}$ in Kaluza-Klein (KK) expansion (\ref{compact}). It is not excluded a priori that non-zero scalar KK modes play a role in inflation and the curvature perturbation. The bare mass of the $(n,m)$ KK mode ($n,m\neq 0$) is given by $m^2_{(n,m)}=(n/R_{5})^2+(m/R_{6})^2\simeq(n/R_5)^2$. On the other hand, the zero mode (the inflaton) mass $m_{\phi}$ arises at one-loop, and is given by $m_{\phi}^2\simeq 4/(\pi^7f^2_5r_{R}R_5^4)\simeq 1/(100r_{R})g^2/R_5^2$, as obtained from (\ref{minf}). Numerically, $m_{\phi}\sim10^{13}\,{\rm GeV}$, by using the values in Table 1. We find that $m_{(n,m)}$ is much larger than $m_{\phi}$. Hence the KK non-zero modes are not slow-rolling fields and will be oscillating. They may have been driven to each minimum before inflation and can be ignored (even if they are oscillating during inflation, they would be diluted away due to inflation).\\
\indent The $(0,1)$ mode need a more careful consideration. In (\ref{g}), there is a parameter region in which the $m_{(0,1)}^2=1/R_6^2$ is smaller than $m_{\phi}^2$ and larger than $m_{\sigma}^2\sim r_{R}g^2/R_6^2$. In this case the (0,1) mode may have some role in inflation and perturbation. A detailed study is necessary to answer to the question of the role of this (0,1) KK mode, and it is a separate work.  \\
\indent It is important to know the reheating temperature in the inflation model which is given by $T_{\rm reh}\sim\sqrt{M_P{\it \Gamma}_{\phi}}$. In our model the inflaton decay width ${\it \Gamma}_{\phi}$ is given by the same form as the curvaton's, ${\it \Gamma}_{\phi}=g^2/(4\pi)m_{\phi}$. Hence, 
\beq
T_{{\rm reh}}\sim\sqrt{\frac{g^2}{4\pi}M_Pm_{\phi}}
                      \sim g\times10^{15}\,{\rm GeV}
                      \sim10^{10}\,{\rm GeV}.\nonumber
\eeq 
The reheating temperature depends only on the size of the gauge coupling constant $g$. \\ 
  \indent  Finally we note that there are many ways that gravity may play roles in the inflation scenarios based on high dimensional theory. One is that the inflaton may arise from the extra components of the higher dimensional metric $g_{MN}$. Then three alternative scenarios should be considered: i) The scalar arising from gauge field $A_M$ dominates the energy density of the universe and plays a role of inflaton such as our model, ii) the scalar from the metric $g_{MN}$ may dominate, iii) the two terms may compete. Which case may realize depends on the model of higher dimensional gauge field/gravity and need a detailed analysis. It will be a separate work and we plan to study this problem in the near future.


\section{Acknowledgments}
We are very grateful to the referee for pointing us a few important questions related to our inflaton model. Replies to these comments helped us clarify the questions and answers. This work is supported partially by the grants for scientific research of the Ministry of Education, Kiban A, 21540278 and Kiban C, 21244063 and by a Chuo University Riko-ken grant. CML was supported partially by the NSC under grant No. NSC 96-2628-M-007-002-MY3, by the NCTS, and by the Boost Program of NTHU.


\end{document}